\documentclass[aps,showpacs,superscriptaddress,twocolumn]{revtex4}
\usepackage{amssymb}
\input epsf
\usepackage{graphicx}
\usepackage{dcolumn}

\begin{document}
\title{Inverted and mirror repeats in model nucleotide sequences}

\author{Fabrizio Lillo}
\affiliation{Dipartimento di Fisica e Tecnologie Relative, Universit\`a di Palermo,
Viale delle Scienze, I-90128, Palermo, Italy}
\affiliation{Santa Fe Institute, 1399 Hyde Park
Road, Santa Fe, NM 87501, USA}

\author{Marco Span\`o}
\affiliation{Dipartimento di Fisica e Tecnologie Relative, Universit\`a di Palermo,
Viale delle Scienze, I-90128, Palermo, Italy}

\begin{abstract}
We analytically and numerically study the probabilistic properties of inverted and mirror repeats in model sequences of nucleic acids. We consider both perfect and non-perfect repeats, i.e. repeats with mismatches and gaps. The considered sequence models are independent identically distributed (i.i.d.) sequences, Markov processes and long range sequences. We show that the number of repeats in correlated sequences is significantly larger than in i.i.d. sequences and that this discrepancy increases exponentially with the repeat length for long range sequences.
\end{abstract} 
\maketitle

\section{Introduction}

The complete sequencing of large genomes has lead to reconsider the importance of non coding DNA or RNA in the regulation of the activity of the cell \cite{Eddy}. Many different types of sequences able to have a regulatory role have been discovered. Among these sequences inverted and mirror repeats play an important role. For example inverted repeats provide the necessary condition for the potential existence of a hairpin structure in the transcribed messenger RNA and/or cruciform structures in DNA \cite{Sinden}. Inverted repeats play also an important role for regulation of transcription and translation. In bacteria, inverted repeats and the associated hairpin structures are often part of rho-independent transcription terminators \cite{Carafa,lesnik}. In recent years there has been a growing interest for these structures triggered by the discovery of new classes of regulatory elements.
Prominent examples of these new regulatory RNA families are microRNA (miRNA) \cite{Lagos,Lau,Lee} and small interference RNA (siRNA) \cite{Hamilton,Hutvagner}. Most of these structures share the property of being associated with an hairpin secondary structure. DNA or RNA short sequences that may be associated to RNA secondary structures are present in genomes of different species of phages, viruses, bacteria and eukaryotes. Indication about the potential existence of RNA secondary structures can be inferred throughout the detection of short pair sequences having the characteristic of inverted repeats in the investigated genomes \cite{schroth,cox,lillo,spano}.   Also mirror repeats may have multiple biological roles. For example, perfect or near-perfect homopurine or homopyrimidine mirror repeats can adopt triple-helical H conformations \cite{mirkin}.  Several computer programs have been developed to detect repeats and/or the associated secondary structure in DNA or RNA sequences \cite{emboss,Warburton}. Few studies have considered the problem of the expected number of repeats in model sequences \cite{Leung,Chew}, mainly investigating the clustering of repeats.

The purpose of this paper is to derive analytical an numerical expressions for the expected number of two specific, yet very important, type of repeats under the assumption that the investigated sequence can be modeled with a given family of stochastic process. In this paper we consider inverted and mirror repeats and we investigate four different types of sequence models. Specifically, we consider independent and identically distributed sequences, first order Markov chains, higher order Markov processes, and long memory sequences. For the first two types of models we are able to derive analytically expressions for the number of repeats, while for the last two classes of models we use numerical simulations to infer phenomenological expressions for the expected number of repeats.  

The outline of the paper is the following. In Section II we introduce the investigated repeats and in Section III we introduce the sequence models discussed in the papers. In Section IV we consider independent and identically distributed sequences and we derive several analytical expressions for repeats. In Section V we consider first order Markov chains and in Section VI we compute numerically the expected number of repeats for higher order Markov processes. In Section VII we consider long memory sequences and Section VIII concludes.

\section{Inverted and Mirror Repeats}

In this paper we consider two types of repeats, i.e. inverted and mirror repeats.
These repeats are composed by two non-overlapping segments of nucleotide sequence that can be separated by another nucleotide subsequence.  A mirror repeat is for example 5'GATTCGAacgAGCTTAG3' where the sequence GATTCGA is repeated in an inverted way after the spacer acg. 
An inverted repeats is for example given by the sequence 5'GATTCGAacgTCGAATC3' where the sequence GATTCGA is repeated and complemented after the spacer acg.
One of the problem in counting repeats is the fact that a single repeat can be counted many times if one does not define in some way a maximal repeat. Consider for example the sequence 5'aggaatcgatcttaacgaagatcgattcca3'. This sequence contains many different inverted repeats, for example, 5'aggAATCGatcttaacgaagatCGATTcca3' or 5'aggaaTCGATCttaacgaaGATCGAttcca3'. If one does not consider inverted with mismatches, there is one {\it maximal} inverted repeats, i.e. 5'aGGAATCGATCTTaacgAAGATCGATTCCa3', in which the first base before and after the structure are not complementary and also the first and the last base of the spacer aacg are not complementary. When one considers inverted or mirror repeats with mismatches the definition of maximal repeat is less clear and must be clearly defined (see Section~\ref{mismatches}). In this paper we are interested in finding the expected number of maximal inverted and mirror repeats in model genome sequences. 

\begin{figure}[ptb]
  \begin{center}
  \includegraphics[scale=0.3]{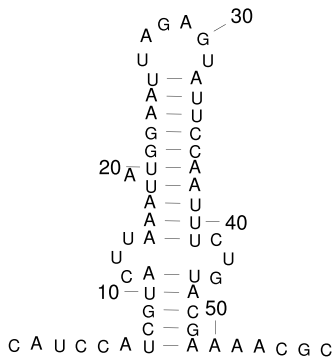} 
  \caption{Secondary structure formed by an inverted repeat in single stranded RNA. Bases from $7$ to $25$ constitute the left arm of the stem and bases from $32$ to $49$ constitute the right arm. The loop is made by bases $26$ to $31$. At base $19$ there is a gap (or one-base bulge), and bases $12$-$14$ and $42$-$44$ constitute a three base mismatch (or internal loop). Note that base $6$ is not complementary to base $50$ and base $26$ is not complementary to base $31$ in order to have maximal repeats.
 According to the terminology used in the paper we have $\ell=18$, $m=6$, $k=3$, $g=1$.}
  \label{inverted}
  \end{center}
\end{figure}

A repeat is characterized by the assignment of a matching rule between couples of nucleotides.  For RNA sequences the matching rule is defined by a $4\times 4$ matrix whose rows and columns correspond to nucleotides A, C, G, and U. A matrix entry is $1$ if the matching between the nucleotides in the row and in the column is allowed and zero elsewhere. For example the characteristic matrix for inverted repeats in which only Watson-Crick base pair (i.e. A-U and C-G) are allowed is 
\begin{eqnarray}
{\bf M}_{(inv)}=\left(\begin{array}{cccc}
 0&0&0&1\\
 0&0&1&0\\
 0&1&0&0\\
 1&0&0&0\\
 \end{array}\right).
\end{eqnarray} 
If the pairing G-U (or GU wobble) is allowed the matrix becomes
\begin{eqnarray}
{\bf M}'_{(inv)}=\left(\begin{array}{cccc}
 0&0&0&1\\
 0&0&1&0\\
 0&1&0&1\\
 1&0&1&0\\
 \end{array}\right).
\end{eqnarray} 
Finally for mirror repeats the characteristic matrix is
\begin{eqnarray}
{\bf M}_{(mir)}=\left(\begin{array}{cccc}
 1&0&0&0\\
 0&1&0&0\\
 0&0&1&0\\
 0&0&0&1\\
 \end{array}\right).
\end{eqnarray} 
Inverted and mirror repeats can be formed both in DNA and in RNA. Since our results are the same for both nucleic acids (provided one replaces T with U), we decide to consider repeats in RNA. 

Given a matching rule, a {\it perfect repeat} of stem length $\ell$ exists at point $x$ when, for a loop value $m$, every base $x+1-i$ matches every base $x+m+i$ for $1\le i \le \ell$. The sequence from $x+1-\ell$ to $x$ will be called left arm of the stem, whereas the sequence from $x+m+1$ to $x+m+\ell$ will be called right arm of the stem. Since we are interested in maximal repeats, the repeats is defined also by requiring that base $x+1$ does not match base $x+m$ and base $x-\ell$ does not match with base $x+m+\ell+1$. We will call these repeats perfect because there are no bulges or mismatches. 
A {\it gap} or {\it one-base bulge} is present in the left arm of the stem if there exists an index $j$ such that the above relation is true for $i\le j$, whereas for $i>j$ every base $x-i$ matches every base $x+m+i$. The extension to bulge in the right arm of the stem is straightforward. Finally, a one nucleotide {\it mismatch} (or internal loop) is present in the stem if for some $i'$ between $1$ and $\ell-2$, the base $x-i'$ does not match with base $x+m+i'+1$. More mismatches or a mismatch composed of more then one base can be present in a stem.   
Inverted repeats are known to be able to create hairpin structures in single strand nucleic acids.
Figure~\ref{inverted} shows an example of hairpin structure formed by an inverted repeat with a bulge and a three base mismatch. The caption should help the reader in understanding the terminology used in this paper.

The purpose of this paper is to derive the expected number of repeats of a given type for simple models of nucleotide sequences. We shall indicate with $N(\ell,m,k,g)$ the expected number of repeats of stem of length $\ell$, loop of length $m$, $k$ one-nucleotide mismatches and $g$ gaps.  The calculation of the expected number of repeats is complex for two main reason. The first problem is to compute the probability $\pi(\ell,m,k,g)$ that a given short sequence generated according to a sequence model can host a repeat with given characteristics. Once this probability is known the next problem is to estimate the expected number of repeats observed in a long sequence (genome) composed by $N$ nucleotides. If the occurrence of different structures were independent one from the other the expected number is simply $N(\ell,m,k,g)=N \pi(\ell,m,k,g)$. Unfortunately, in general the occurrence of a given structure is not independent of the presence of another structure. In the statistical search of simple words in genomes this is a known problem (see for example \cite{waterman}). However, since we search for maximal repeats and the structure we are interested in are long and complex, we neglect the problem of non independence. In all the cases considered below we have performed extensive numerical simulations to test our formulas and, indirectly, the independence assumption. By performing careful statistical tests (usually $\chi^2$ tests) we cannot reject the hypothesis that our formulas are correct. For this reason in the following we present the formulas for $N(\ell,m,k,g)$ rather than for  $\pi(\ell,m,k,g)$.

\section{Models for nucleotide sequences} 

\subsection{Independent Identically Distributed Sequences}

The simplest model for nucleotide sequences is the independent identically distributed (i.i.d.) model. In this model one assumes independent nucleotides with probabilities $p_a$, $p_c$, $p_g$, and $p_u$, such that $p_a+p_c+p_g+p_u=1$. Although it is known that correlation between nucleotides are significant, this model allows exact analytical calculations and can be used as a useful starting point.

It is useful to define the probability vector ${\bf p}^T\equiv (p_a,p_c,p_g,p_u)$ where the elements are the nucleotide probabilities. Given a type of structures characterized by the matrix ${\bf M}$ we introduce the scalar quantity
\begin{equation}
q={\bf p}^T {\bf M} {\bf p}.
\end{equation}
For example, inverted repeats have $q=2p_ap_u+2p_cp_g$, whereas for mirror repeats $q=p_a^2+p_c^2+p_g^2+p_u^2$.

\subsection{Markov models}
A better class of models for nucleotide sequences is the class of Markov processes. Let us consider for convenience the  infinite sequence $X_i$, where $i\in \mathbb{Z}$ and  $\mathbb{Z}$ is the set of integers. An ergodic stationary $m$-th order Markov chain is characterized by the transition matrix
\begin{eqnarray}
p(a_{m+1}|a_1,....,a_m)\\ \nonumber
=P(X_i=a_{m+1}|X_{i-1}=a_m,...,X_{i-m}=a_1).
\end{eqnarray}

The simplest Markov chain we shall consider extensively in the following is the $1$-st order Markov chain. This type of processes is characterized by the $4\times 4$ transition matrix $p(a_2|a_1)$. By taking powers of this matrix one can also define the $k$-step transition matrix whose elements are
$p_k(b|a)=P(X_i=b|X_{i-k}=a)$. In this notation $p(a_2|a_1)=p_1(a_2|a_1)$.

The model parameters, i.e. the order of the Markov chain and the transition probabilities, of a real sequence can be estimated by the maximum-likelihood method (see for example \cite{waterman}).

\subsection{Long memory models}

In recent years it has been proposed that parts of real genomes are not well described by Markovian models, but rather that a long memory (or long-range) process describes better the correlation properties of nucleotide sequences \cite{Peng92,Li92,voss, Mantegna94,Buldyrev95}. There are several ways of detecting and modeling correlation properties of nucleotide sequences. The approach we will follow is called ``DNA walk" \cite{Peng92} and consists in mapping the nucleotide sequence in a one-dimensional random walk $x$. Since there are $4$ different residues in a RNA sequence while the random walk has two possible directions ($\Delta x=\pm1$), one needs to choose a mapping rule from the $4$ residues to the $2$ directions. Several different mapping rules have been introduced \cite{Buldyrev95}. In the present paper we consider two important rules: (i) the purine-pyrimidine rule (or RY rule) which assigns $\Delta x=+1$ if the residue is a purine (A or G) and $\Delta x=-1$ if the residue is a pyrimidine (C or U) and (ii) the hydrogen bond energy rule (or SW rule) which assigns  $\Delta x=+1$ for strongly bonded residues (C or G) and assigns $\Delta x=-1$ for weakly bonded residues (A or U). This second rules can be useful to take into accounts the isochore structure of genome \cite{bernardi}. By using either of these rules it has been observed that in most cases non-coding DNA sequences, i.e. DNA sequences not coding for proteins, display long-memory properties of the corresponding DNA walk. We remind that a long-memory process is a process whose autocorrelation function of $\Delta x_i$ decays in time as $Corr[\Delta x_{i+\tau} \Delta x_i]\sim \tau^{-\gamma}$, where $0<\gamma<1$. Long memory processes are an important class of stochastic process that have found application in many different fields \cite{Beran94}. The autocorrelation function of a long memory process is not integrable in  $\tau$ between $0$ and $+\infty$ and, as a consequence, the process does not have a typical time scale. Long memory processes are better characterized by the Hurst exponent $H$ that, for long memory processes, is $H=1-\gamma/2$.  Thus for long-memory processes $1/2<H<1$.

Long memory properties of nucleotide sequences has been associated to different genome characteristics including nucleosomal structure in eukaryotes \cite{audit}, to the presence of isochores \cite{bernardi} and to the presence of tandem repeats \cite{holste}. More recently it has been suggested that in some genomes (for example, human) the correlation properties of DNA cannot be captured by a single Hurst exponent, but rather that the Hurst exponent may depend on the observation scale \cite{liholste,carpena}. Different scales can be associated with different biological structures (genes, transposable elements, isochores).

\section{Inverted and mirror repeats in IID sequences}

\subsection{Perfect repeats}
The expected number of perfect repeats of stem length $\ell$ and loop length $m$ in a i.i.d. genome of length $N$ characterized by the parameter $q$ is
\begin{equation}
N(\ell,m)=N(1-q)^{\alpha}q^{\ell},
\label{basic}
\end{equation}  
where the exponent $\alpha$ is equal to $1$ for $m\le1$ and is equal to $2$ for $m\ge2$. In other words we need to impose that the $\ell$ bases of the left arm of the stem match with the corresponding bases in the right arm. Moreover we need to impose that the  first couple of bases in the loop does not match, such as the first couple of bases at the end of the stem. When the loop is shorter that $2$ nucleotides one cannot impose that the first couple of bases in the loop does not match and this explains the different value of the exponent $\alpha$. Since in a i.i.d. sequence the occurrences of nucleotide are independent probabilities factorize and Eq.~\ref{basic} is obtained.  This expression has been used, for example, in Ref.~\cite{lillo} to investigate the number of perfect inverted repeats in bacterial genomes.

\subsection{Inverted with mismatches}\label{mismatches}

A mismatch in a repeat is the presence of a pair of nucleotides in the stem that do not match. We indicate with $k$ the number of mismatches in the stem and we look for an expression for $N(\ell,m,k)$.
We prove that the expected number is
 \begin{equation}
N(\ell,m,k)=N {\ell-2 \choose k} (1-q)^{\alpha+k}q^{\ell-k},
\label{mism}
\end{equation}
where the exponent $\alpha$ assumes the same values as in Eq.~\ref{basic}. In fact a mismatch can be present only in one of the $\ell-2$ internal nucleotide of the stem (i.e. from the second to the $(\ell-1)$-th nucleotide). There are ${\ell-2 \choose k}$ ways of placing $k$ mismatches in $\ell-2$ internal bases of the stem.

One of the problem of Eq.~\ref{mism} is the fact that, for example, a  repeat with one mismatch can also be seen as a repeat with zero mismatches and a shorter stem. We shall denote these two repeats as {\it embedded}. One is usually interested in counting more embedded  repeats only once. Moreover programs designed for the search of inverted repeats, such as {\it palindrome} of the EMBOSS package \cite{emboss}, effectively count embedded inverted repeats only once. Therefore we need a formula for non embedded  repeats. Clearly any  repeat with, say, zero mismatches can be thought as part of a longer  repeat with a large number of mismatches. In other words we need to introduce an upper value of the number of mismatches, in order to find an expression for non embedded repeats up to a chosen value of the number of possible mismatches. For example we can ask for the expected number of  inverted repeats with zero mismatches that cannot be seen as part of longer inverted repeats with one mismatch. This of course does not guarantee that the found  repeats cannot be part of  repeats with two mismatches. From an operative point of view, this corresponds to run the search program (for example {\it palindrome}) with a maximal number of mismatches equal to $\bar k$. Therefore a quantity more meaningful than Eq.(\ref{mism}) is $N^{(\bar k)}(\ell,m,k)$, which is the expected number of  repeats of stem length $\ell$, loop length $m$, and $k$ mismatches, that cannot be part of a longer  repeat of the same type with at most $\bar k$ mismatches. By definition $\bar k\ge k$.
The two expressions of Eq.s~\ref{basic} and \ref{mism} correspond to  $N^{(0)}(\ell,m,0)$ and  $N^{(k)}(\ell,m,k)$, respectively.

When $\bar k =1$ we have
\begin{eqnarray}
 N^{(1)}(\ell,m,0)=N(1-q)^{\alpha}q^{\ell} \nonumber\\ 
 \alpha=\left \{\begin{array}{ll}
                                                   ~~2 & {\rm{for}} ~0\le m \le 1 \\
                                                   ~~3 & {\rm{for}} ~2\le m\le 3  \\
                                                   ~~4 & {\rm{for}} ~m\ge 4.
                                    \end{array} \right.               
\end{eqnarray}

When $\bar k=2$ we have 
\begin{eqnarray}
 N^{(2)}(\ell,m,1)= N (\ell-2) (1-q)^{\alpha+1}q^{\ell-1}\nonumber\\ 
 \alpha=\left \{\begin{array}{ll}
                                                   ~~2 & {\rm{for}} ~0\le m \le 1 \\
                                                   ~~3 & {\rm{for}} ~2\le m\le 3  \\
                                                   ~~4 & {\rm{for}} ~m\ge 4.
                                    \end{array} \right.               
\end{eqnarray}
and 
\begin{eqnarray}
 N^{(2)}(\ell,m,0)=N(1-q)^{\alpha}q^{\ell} \nonumber\\ 
 \alpha=\left \{\begin{array}{ll}
                                                   ~~3 & {\rm{for}} ~0\le m \le 1 \\
                                                   ~~4 & {\rm{for}} ~2\le m\le 3  \\
                                                   ~~5 & {\rm{for}} ~4\le m\le 5 \\
                                                   ~~6 & {\rm{for}} ~m\ge 6.
                                    \end{array} \right.               
\end{eqnarray}

The general formula is
\begin{eqnarray}
 N^{(\bar k)}(\ell,m,k)=N  {\ell-2 \choose k} (1-q)^{\alpha+\beta}q^{\ell-k}\nonumber\\ 
  \alpha=\left \{\begin{array}{ll}
                                                   ~~1 & {\rm{for}} ~0\le m \le 1 \\
                                                   ~~2 & {\rm{for}} ~m\ge 2
                                    \end{array} \right.    \nonumber\\
 \beta=(\bar k -k)+\max\left(0,\min(\left[\frac{m}{2}\right]-1, \bar k-k)\right),  
 \label{general_mismatch}             
\end{eqnarray}     
where $[x]$ indicates the integer part of $x$.  

We have performed extensive numerical simulations of artificial genomes and we have verified that these expressions are correct. Specifically, we have written computer programs able to detect inverted or mirror repeats with the required characteristics (stem and loop length, mismatches, etc.). Then we have performed a $\chi^2$ test between the frequency of observed repeats and the frequency expected by our theory. In all cases we cannot reject the hypothesis that our formulas are correct.

\subsection{Repeats with one gap}

\begin{figure}[ptb]
  \begin{center}
  \includegraphics[scale=0.33]{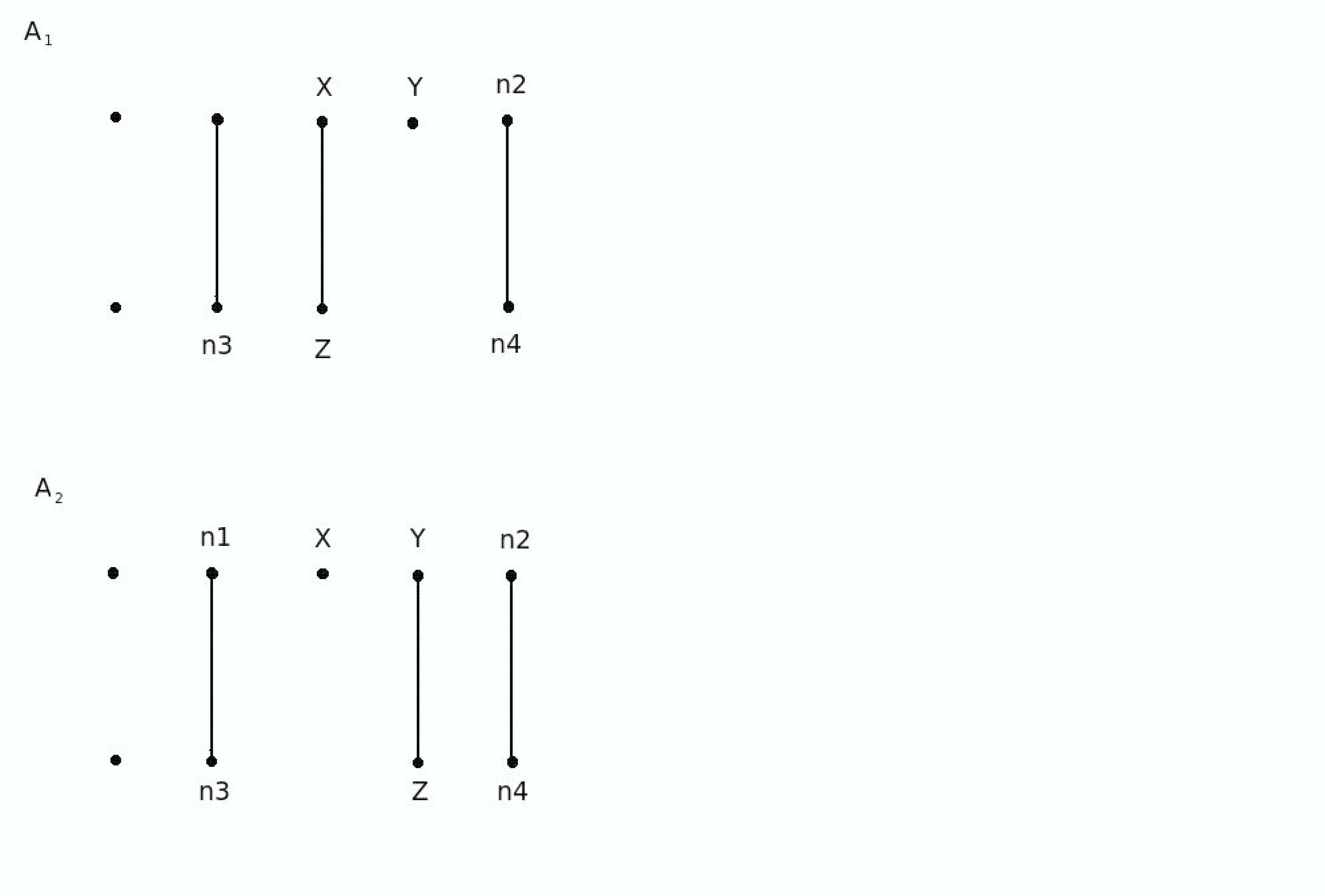} 
  \caption{Schematic representation of the stem of the two possible secondary structures formed by an inverted repeat with stem length $\ell=3$ and one gap. The continuous lines indicate complementarity and the labels on the bases are used in the text.}
  \label{gapfig}
  \end{center}
\end{figure}

We consider now  the case of inverted and mirror repeats with one gap in the stem and no mismatches. We shall indicate with $\ell$ the number of links in the stem, since in such a structure there will be $\ell$ nucleotides in one branch of the stem and $\ell+1$ in the other. The expected number of repeats with the gap in one {\it specific} position is the same as for perfect repeats (see Eq.~\ref{basic}), i.e. 
\begin{equation}
N(\ell,m,k=0,g=1)=N(1-q)^{\alpha}q^{\ell},
\label{basic_gap}
\end{equation}  
where the exponent $\alpha$ is equal to $1$ for $m\le1$ and is equal to $2$ for $m\ge2$. One could think that, since there are $\ell-1$ possible positions for the gap (on one arm), the expected number of repeats with one gap in any position of one arm is simply $\ell-1$ times the value in Eq.~\ref{basic_gap}. This is wrong because the probability of observing the gap in one position is not independent from the probability of observing the gap in another position. To understand why, let us consider an inverted repeat with $\ell=3$ and one gap. As shown in Fig.~\ref{gapfig} there are two positions for the gap, and the corresponding structures are indicated as $A_1$ and $A_2$ in the figure.   
The probability of observing either $A_1$ or $A_2$ or both is
\begin{equation}
P(A_1\cup A_2)=P(A_1)+P(A_2)-P(A_1\cap A_2).
\end{equation}
$P(A_1)$ and $P(A_2)$ are equal to the quantity in Eq~\ref{basic_gap}, whereas $P(A_1\cap A_2)$ is the joint probability that the sequence can form both structures $A_1$ and $A_2$.  By looking at the figure we note that the sequence can form both structures if $X=Y=\bar Z$, where the bar indicates complementarity. Thus the joint probability is
\begin{eqnarray}
P(A_1\cap A_2)=  (1-q)^{\alpha}q^{2} (p_a^{2}p_t+p_ap_t^{2}+p_c^{2}p_g+p_cp_g^{2})\nonumber \\
 \equiv (1-q)^{\alpha}q^{2} \tilde q.~~~~~~~~~~~ 
\end{eqnarray}
For inverted repeats the quantity $\tilde q$ is the probability that $X=Y=\bar Z$ and it is equal to $p_a^{2}p_t+p_ap_t^{2}+p_c^{2}p_g+p_cp_g^{2}$. Analogously for mirror repeats $\tilde q$ is the probability that $X=Y=Z$ and it is equal to $p_a^3+p_t^3+p_c^3+p_t^3$.   In conclusion, the expected number of repeats with $\ell=3$ and one gap is
\begin{equation}
N(\ell=3,m,k=0,g=1)=N(1-q)^{\alpha}q^{2}(2q-\tilde q),
\label{gap3}
\end{equation}  
which is of course different from the naive (and wrong) answer given by twice Eq.~\ref{basic_gap}. The generalization of this last formula to a generic value of $\ell$ is not straightforward and the derivation is reported in Appendix 1. The result is
\begin{eqnarray}
 N(\ell,m,k=0,1)=2Nq^{\ell-1}(1-q)^{\alpha} [(\ell-1)q-(\ell-2)\tilde q]               \nonumber\\ 
 \alpha=\left \{\begin{array}{ll}
                                                   ~~1 & {\rm{for}} ~0\le m \le 1 \\
                                                   ~~2 & { \rm{for}} ~m\ge 2,  
                                    \end{array} \right.               
\label{IRswithgap2}
\end{eqnarray}
where the factor $2$ in front $q^{\ell-1}$ is due to fact that the gap can be found in one of the two arms. It is worth noting that for large $\ell$ the correct answer of Eq.~(\ref{IRswithgap2}) is $3/4$ of the naive and wrong answer given by $\ell-1$ times the expression of Eq.~(\ref{basic_gap}).

\section{Inverted and mirror repeats in first order Markov chains}

We now give the expression for the expected number of  repeats for a model sequence described by a $1$-st order Markov chain. We consider the simpler case of the expected number of perfect  repeats with a given stem (of length $\ell$, as before) and a generic loop of length $m>2$.

The calculation is performed in Appendix 2 and the result is
\begin{widetext}
\begin{eqnarray}
P_{markov}(\ell,m)=\sum_{n_1,n_2,,...,n_{\ell}=1}^4 p(n_1n_2...n_{\ell})p(\bar n_{\ell}...\bar n_2\bar n_1)\nonumber\\
\times\frac{\left(p(n_1)-\sum_{x=1}^4p(n_1|x)p(\bar x|\bar n_1)\right) \left(p_{m+1}(\bar n_{\ell}|n_{\ell})-\sum_{y=1}^4 p(\bar n_{\ell}|y)p_{m-1}(y|\bar y)p(\bar y|n_{\ell})\right)}{p(n_1)p(\bar n_{\ell})},
\label{markov_general}
\end{eqnarray}
\end{widetext}
where $\bar n_i$ indicates a base matching with base $n_i$, i.e. the complementary of $n_i$ for inverted repeats and $\bar n_i=n_i$ for mirror repeats. In Eq.~\ref{markov_general}  $p(n_i)$ is the probability of occurrence of base $i$ and $p(n_1n_2...n_{\ell})$ is the probability of occurrence of the word $n_1n_2...n_\ell$, that for Markov chain is easily computable (see also Appendix 2). 
Even if the expression (\ref{markov_general}) looks complex, the numerical summation is easily and quickly performed for example with simple programs in Mathematica. It is worth noting that the summation is over $4^\ell$ terms, whereas a direct calculation taking into account all the possible repeats would require to sum $4^{2\ell+2+m}$ terms.

The functional dependence of $P_{markov}(\ell,m)$ from $\ell$ and $m$ are not evident by eye, such as the relative magnitude of $P_{markov}(\ell,m)$ and $P_{iid}(\ell,m)=(1-q)^{\alpha}q^{\ell}$ for an i.i.d genome (see Eq.(\ref{basic})). Thus we discuss here these issues by considering Markov models with parameters equal to the ones obtained by real genomes of model organisms. Specifically we shall consider four complete genomes: (i) the Hepatitis B virus (accession NC\_003977, length$=3,215$ bp), (ii) the {\it Escherichia coli} K12 genome (accession NC\_000913, length=$ 4,639,675$ bp), (iii) the {\it Drosophila melanogaster} mitochondrion (accession NC\_001709, length=$19,517$ bp), and (iv) the {\it Homo sapiens} mitochondrion (accession NC\_001807, length=$16,571$ bp). Moreover we consider inverted repeats.

\begin{figure}[ptb]
  \begin{center}
  \includegraphics[scale=0.33,angle=-90]{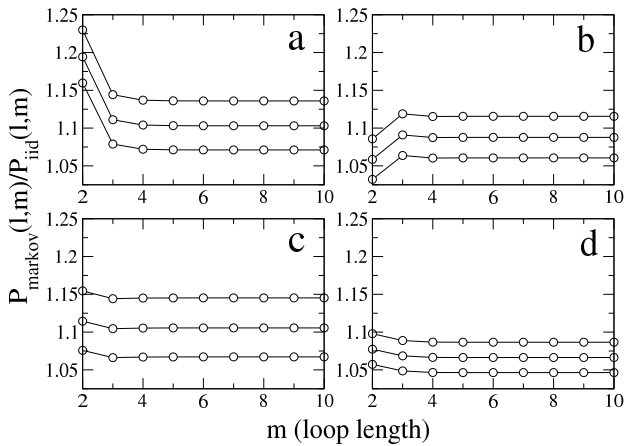} 
  \caption{Plots of the ratio $P_{markov}(\ell,m)/P_{iid}(\ell,m)$ between the probability of observing an inverted repeat with stem length $\ell$ and loop length $m$ in a Markov and in an i.i.d. genome as a function of the loop length $m$. The parameters characterizing the models are estimated by four model genomes, i.e. Hepatitis B virus (a), {\it E. coli} (b), {\it Drosophila} mitochondrion (c), and {\it Homo} mitochondrion (d). In each panel the curves refer to $\ell=4$, $\ell=5$, and $\ell=6$ (from bottom to top).}
  \label{m_dependence}
  \end{center}
\end{figure}

We first discuss the dependence of $P_{markov}(\ell,m)$ from the loop length $m$. To this end we computed the ratio $P_{markov}(\ell,m)/P_{iid}(\ell,m)$ for the stem length fixed at $\ell=4,5,$ and $6$.
Figure~\ref{m_dependence} shows this quantity for the four model genomes. We see that  $P_{markov}(\ell,m)$ has a small dependence from $m$. More precisely for $m$ larger than few units, $P_{markov}(\ell,m)/P_{iid}(\ell,m)$ becomes independent on $m$. The loop length dependence for small values of $m$ can be positive (panels a,c, and d) or negative (panel b) with respect to the value for large $m$. In all cases the ratio $P_{markov}(\ell,m)/P_{iid}(\ell,m)$ is significantly larger than one and it increases with the stem length $\ell$. 

Because of the small dependence on $m$ we can consider $P_{markov}(\ell,m)$ for large values of $m$ as a good approximation of the probability of observing  repeats. This approximation leads to a simplification of Eq.~\ref{markov_general}. In fact, when $m$ is large one can approximate the conditional probabilities in Eq.~\ref{markov_general} $p_{m+1}(\bar n_{\ell}|n_{\ell})\simeq p(\bar n_{\ell})$ and $p_{m-1}(y|\bar y)\simeq p(y)$. Thus the probability $P_{markov}(\ell,m)$ becomes independent from $m$ and equal to 
\begin{widetext}
\begin{eqnarray}
P_{markov}(\ell,m)=\sum_{n_1,n_2,,...,n_{\ell}=1}^4 p(n_1n_2...n_{\ell})p(\bar n_{\ell}...\bar n_2\bar n_1)\\
\times\frac{\left(p(n_1)-\sum_{x=1}^4p(n_1|x)p(\bar x|\bar n_1)\right) \left(p(\bar n_{\ell})-\sum_{y=1}^4 p(\bar n_{\ell}|y)p(y)p(\bar y|n_{\ell})\right)}{p(n_1)p(\bar n_{\ell})},\nonumber
\label{markov_asymptotic}
\end{eqnarray}
\end{widetext}

\begin{figure}[ptb]
  \begin{center}
  \includegraphics[scale=0.33,angle=-90]{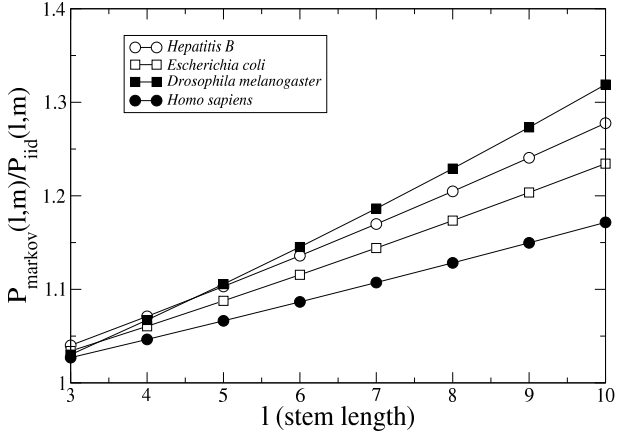} 
  \caption{Plots of the ratio $P_{markov}(\ell,m)/P_{iid}(\ell,m)$ between the probability of observing an inverted repeat with stem length $\ell$ and loop length $m>5$ in a Markov and in an i.i.d. genome as a function of the stem length $\ell$. The parameters characterizing the models are estimated by four model genomes, i.e. Hepatitis B virus (empty circles), {\it E. coli} (empty squares), {\it Drosophila} mitochondrion (filled squares), and {\it Homo} mitochondrion (filled circles).}
  \label{l_dependence}
  \end{center}
\end{figure}

We can now study the dependence of $P_{markov}(\ell,m)$ from the stem length $\ell$, by considering the cases when $m$ is larger than $4$ bp. Figure~\ref{l_dependence} shows the ratio $P_{markov}(\ell,m)/P_{iid}(\ell,m)$ as a function of $\ell$ for the four genomes. In all cases the ratio $P_{markov}(\ell,m)/P_{iid}(\ell,m)$ increases almost linearly with the stem length $\ell$. For $\ell\le 10$ the order of magnitude of the error made by the iid model in predicting the number of repeats of a Markov sequence ranges between few percents and $30\%$.

\subsection{A simplified model}\label{simplified_sec}
The fact that even for large values of $m$ the number of inverted repeats expected in a Markovian genome is significantly larger than the number expected in an iid genome can be explained in a simplified model of genome sequence. We assume that the nucleotide alphabet is composed only by two symbols (instead of four), that the transition matrix is parameterized as
\begin{eqnarray}
\left(\begin{array}{cc}
 \frac{1}{2}+\delta&\frac{1}{2}-\delta\\
 \frac{1}{2}-\delta&\frac{1}{2}+\delta\\
 \end{array}\right), 
 \end{eqnarray}
 and that the process is stationary, so that the probability for the two symbols are equal to $1/2$. The parameter $\delta$ is a measure of the distance from the iid model. With this transition matrix, the conditional probability $p(n_2|n_1)$ is equal to $1/2+\delta$ if $n_1=n_2$ and to $1/2-\delta$ if $n_1\ne n_2$. We shall call permanence the first case and change the second one. We simplify further the original model by removing the constraints that the repeat is maximal, i.e. the condition that the two bases before and after the repeat are not complementary and that the first and last base in the loop are not complementary. The probability of an inverted repeat of stem length $\ell$ and loop length $m>>1$ is given by the product of the probability of the left part of the stem times probability of the right part of the stem. The probabilities factorize because we have assumed that the loop is large. Now the probability for a given word in the left part of the stem is $2^{-1}(1/2-\delta)^{d_1}(1/2+\delta)^{d_2}$, where $d_1$ is the number permanencies, whereas $d_2$ is the number of changes. Clearly it is $d_1+d_2=\ell-1$. The probability for the inverted and complemented word in the right arm of the stem is equal, so the probability for a given inverted is $[2^{-1}(1/2-\delta)^{d_1}(1/2+\delta)^{d_2}]^2$. We have to sum this quantity over all possible words, i.e.
 \begin{eqnarray}
 P(\ell)=\frac{2}{4}\sum_{d_1=0}^{\ell-1}{\ell-1 \choose d_1}\left(\frac{1}{2}+\delta\right)^{2d_1}\left(\frac{1}{2}-\delta\right)^{2(\ell-1-d_1)}\nonumber \\=\frac{1}{2}\left(\frac{1}{2}+2\delta^2\right)^{\ell-1},~~~~~~~
 \end{eqnarray}   
where the factor $2$ in front of the sum comes from the fact there are two possible words with the same position of the permanencies  and of the changes obtained by exchanging one symbol with the other. 
 For an iid sequence the probability for an inverted of stem length $\ell$ is $P_{iid}(\ell)=2^{-\ell}$, thus the ratio is
\begin{equation}
\frac{P(\ell)}{P_{iid}(\ell)}=\frac{\frac{1}{2}\left[\frac{1}{2}+2\delta^2\right]^{\ell-1}}{\frac{1}{2^{\ell}}}=(1+4\delta^2)^{\ell-1},
\label{simplified}
\end{equation}
For small values of $\delta$, i.e. for Markovian sequences not too different from iid ones, the binomial expansion gives
\begin{equation}
 \frac{P(\ell)}{P_{iid}(\ell)}\simeq 1+(\ell-1)4\delta^2,~~~~~~~~~~~~~~~\delta<<1,
 \end{equation}
which is the almost linear behavior observed in Figure~\ref{l_dependence}. Thus we expect the linear behavior observed in Fig.~\ref{l_dependence} for the more complete model is valid for moderate value of the stem. 
 
\section{Higher order Markov models}

\begin{figure}[ptb]
  \begin{center}
  \includegraphics[scale=0.33,angle=-90]{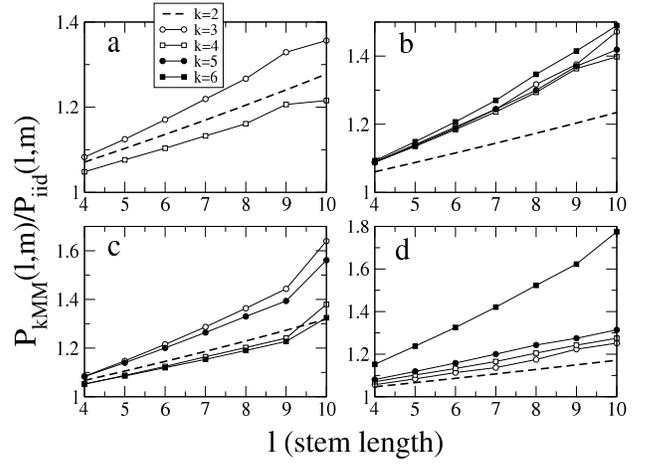} 
  \caption{Plots of the ratio $P_{kMM}(\ell,m)/P_{iid}(\ell,m)$ between the probability of observing an inverted repeat with stem length $\ell$ and loop length $m>5$ in a $k-$th order Markov and in an i.i.d. genome as a function of the stem length $\ell$. The parameters characterizing the models are estimated by four model genomes, i.e. Hepatitis B virus (empty circles), {\it E. coli} (empty squares), {\it Drosophila} mitochondrion (filled squares), and {\it Homo} mitochondrion (filled circles).}
  \label{highorder}
  \end{center}
\end{figure}

In the case of higher order Markov processes the analytical computation of the expected number of inverted and mirror repeats becomes considerably more complex. Instead of trying to obtain complicated expression with difficult interpretation, we perform numerical simulations of higher order Markov chains and we compare the observed number of repeats with the number expected from the iid theory.
The results of our simulations are shown in fig.~\ref{highorder} and indiate that the error made in using an iid model to estimate the expected number of inverted repeats in a Markov chain increases with (i) the stem length $\ell$ and (ii) the order of the Markov process. Nevertheless it is worth pointing out that for moderate values of the stem length the ratio $P_{kMM}(\ell,m)/P_{iid}(\ell,m)$ increases approximately linearly with $\ell$. This implies that in the considered range the number of inverted repeats in a Markovian genome is given by
\begin{equation}
P_{kMM}(\ell,m)\sim A_k \ell ~q^{\ell},
\end{equation}
where $A_k$ is a parameter which slowly increases with the order $k$ of the Markov process.

\section{Long memory processes}

\begin{figure}[ptb]
  \begin{center}
  \includegraphics[scale=0.33,angle=-90]{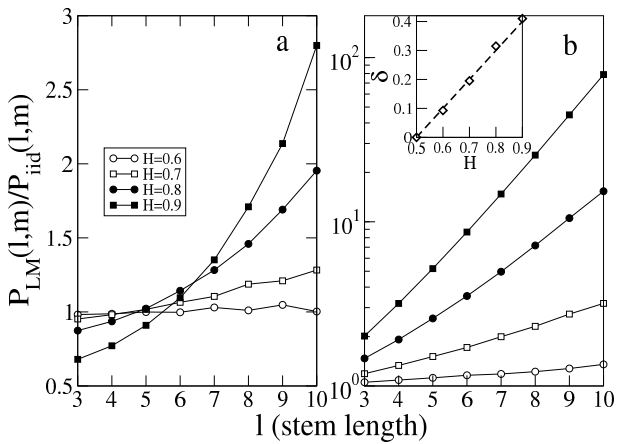} 
  \caption{Plots of the ratio $P_{LM}(\ell,m)/P_{iid}(\ell,m)$ between the probability of observing an inverted repeat with stem length $\ell$ and loop length $m>5$ in a long memory and in an i.i.d. genome as a function of the stem length $\ell$ and of the Hurst exponent $H$. Panel (a) shows the RY (or purine-pyrimidine) rule and panel (b) shows the SW (or hydrogen bond energy) rule. The inset of panel (b) shows the fitted $\delta$ (see text) as a function of $H$. The dashed line is the function $\delta=H-1/2$. For each value of $H$ we simulated an artificial genome of length $10^8$ bp.}
  \label{longmemory}
  \end{center}
\end{figure}

Finally we consider the problem of estimating numerically the probability of occurrence of an inverted or a mirror repeat in a long-range nucleotide sequence. Since most of the repeats with biological role are likely to be find in non-coding regions of the genome which are often composed by long memory nucleotide sequences, this analysis is particular relevant for application to real cases. We generated long memory nucleotide sequences by using either the RY rule or the SW rule and with different values of the Hurst exponent $H$. For example to generate a RY long memory genome we simulated a binary long memory process with values $x_i=\pm1$. Then for each $x_i=+1$ we associated either a A or a G each with probability $1/2$ and for each  $x_i=-1$ we associated either a C or a U each with probability $1/2$. Note that with this generation algorithm the simulated genomes have equal nucleotide frequencies, i.e. $p_a=p_c=p_g=p_t=1/4$. We then searched in the simulated genome for perfect repeats with a given stem length $\ell$ and loop length $m$ and we compare the observed frequencies with the one expected by an iid genome. First of all we find that also for long memory sequences the occurrence of inverted or mirror repeats is essentially independent on the value of the loop length $m$.  As for the Markovian case we find a small dependence for very small values of $m$. The behavior as a function of the stem length $\ell$ is very different from the iid case. In figure~\ref{longmemory} we plot the quantity  $P_{LM}(\ell,m)/P_{iid}(\ell,m)$ as a function of $\ell$, where  $P_{LM}(\ell,m)$ is the observed probability of inverted repeats in the long memory sequence. The left panel shows the RY (or purine-pyrimidine) rule and the right panel shows the SW (or hydrogen bond energy) rule. In the RY case for $\ell\lesssim 5$ there is a decrease of the number of inverted repeats with respect to the iid case whereas for $\ell\gtrsim5$ the number of observed inverted repeats is larger than the number expected in the iid case. However the value of the ratio $P_{LM}(\ell,m)/P_{iid}(\ell,m)$ is never vary large. For the SW rule a different behavior is observed. In right panel of fig.~\ref{longmemory} the y axis is in a logarithmic scale and the ratio $P_{LM}(\ell,m)/P_{iid}(\ell,m)$ has a clear exponential dependence on $\ell$. Very large value of the ratio are observed showing that using the iid formula for long memory sequence can lead to a severe underestimation of the expected repeats. The difference observed between the two rules can be easily explained by recalling that an inverted repeats is formed when many bonds  can be formed between complementary bases. Since in the SW rule the presence of, say, a C is strongly correlated with the presence of a G nearby, it is intuitive to understand why many more inverted repeats are observed in a SW than in a RY long memory genome with the same Hurst exponent.    

Since it is difficult to develop a theory for the number of repeats in a long memory genome, we try to get some intuition by considering the simplified model for Markovian genomes presented in section~\ref{simplified_sec}. We remind that Eq.~\ref{simplified} predicts that the ratio $P(\ell)/P_{iid}(\ell)$ depends exponentially from $\ell$ according to $\exp(\ell~\ln(1+4\delta^2))$ where $\delta$ quantifies the ``distance" of the model from the iid case. We fitted the curves in the right panel of fig~\ref{longmemory} with an exponential function and we estimated the corresponding value of $\delta$ as a function of $H$. The inset of the right panel of figure~\ref{longmemory} shows that to a good approximation $\delta=H-1/2$. This allows us to conjecture that the number of inverted repeats in SW long memory sequences is 
\begin{eqnarray}
P_{LM}(\ell,m)=P_{iid}(\ell,m)~\exp[\ell(1+4(H-1/2)^2)] \nonumber \\
\simeq q^\ell~\exp[\ell(1+4(H-1/2)^2)].
\label{longmemoryeq}
\end{eqnarray}  
For mirror repeats we find that long memory sequences generated according to either SW or the RY rule show a behavior essentially indistinguishable from the one shown in the right panel of Fig.~\ref{longmemory}. The reason is that both rules significantly increase the probability that two equal symbols are found at a short distance. As a consequence Eq.~\ref{longmemoryeq} holds also for mirror repeats according to either SW or RY rule.  
We stress again that this formula holds for sequences with approximately equal nucleotide frequencies.
In conclusion, differently from the Markov case, the exponential behavior of $P(\ell)/P_{iid}(\ell)$ expected from the simplified model is observable in long memory sequences also for small values of $\ell$. This is very important because it means that when the sequence is long memory (as in many non coding sequences) the expected number of repeats can be significantly larger than the number expected in an iid sequence. The discrepancy between iid and long memory models increases very quickly with $H-1/2$. Many regions of real genomes can have very large values of $H$. For example, parts of the human chromosome $22$ have an estimated Hurst exponent $H=0.88$ \cite{bernaola}. In these cases a careful modeling of the nucleotide sequence is very important in estimating the expected number of repeats.

\section{Conclusions}

In conclusion we have developed many analytical and numerical results for the expected number of inverted and mirror repeats with different features (stem length, loop length, presence of mismatches or gap) under the assumption that the investigated sequence can be modeled with different types of sequence models. In general the computation of the number of repeats in model sequences is a complicated problems due to combinatorial difficulties, non independence of different occurrences (as in the case of gaps), and difficulties related to the sequence model (as for higher order Markov process and long memory sequences). To the best of our knowledge this is the most comprehensive study of the occurrence of inverted and mirror repeats in model sequences. A careful estimation of the expected number of repeats in a model sequence is crucial when the investigation of a real sequence displays the presence of an high number of repeats. Is this high number expected under some realistic hypothesis of the sequence model? Without a clear answer to this question it is very difficult to assess if the number of repeats observed in the real sequence has a potential biological role because the repeats are over-represented. The set of results we have obtained in this paper could usefully complement the repeat search algorithms to give a measure of the significance of the number detected occurrences.  

\acknowledgments

We wish to thank Rosario Mantegna and Salvatore Miccich\'e for useful discussions. We acknowledge financial support from the NEST-DYSONET 12911 EU project.

\section{Appendix 1}

In this appendix we derive Eq.~\ref{IRswithgap2} for the number of repeats with stem length $\ell$ and  one gap. 

There are $\ell-1$ possible positions for the gap in one arm. Let us call $A_i$, ($i=1,..,\ell-1$) the set of structures in which the gap has the $i$-th position (see Fig.~\ref{IRswithgap2} for the case $\ell=3$). This ensemble of sets has the property that for any set of indices $i_1<i_2<...<i_k$ it is
\begin{equation}
P(A_{i_1}\cap A_{i_2}\cap...\cap A_{i_k})=P(A_{i_1}\cap A_{i_k}).
\label{setproperty}
\end{equation}
In fact if the sequence under consideration can form a structure with the gap both in the $i_1$ and the $i_k$ position, then it can form the structure with the gap in any intermediate position.

We state the following theorem.

{\bf Theorem} Given an ensemble of sets $A_1,A_2,....,A_N$ satisfying the property (\ref{setproperty}), it holds
\begin{equation}
P(A_1\cup A_2 \cup....\cup A_N) =\sum_{i=1}^NP(A_i)-\sum_{i=1}^{N-1} P(A_i\cap A_{i+1}).
\label{theorem}
\end{equation}
In order to prove this theorem we need a lemma.

{\bf Lemma} Under the above hypothesis (\ref{setproperty}), it is
\begin{equation}
P[\cup_{i=1}^n (A_i\cap A_{n+1})]=P(A_n\cap A_{n+1}).
\end{equation}
In fact
\begin{widetext}
\begin{eqnarray}
P[\cup_{i=1}^n (A_i\cap A_{n+1})]=P\lbrace [\cup_{i=1}^{n-1}(A_i\cap A_{n+1})]\cup [A_n\cap A_{n+1}]\rbrace\nonumber \\
=P[[\cup_{i=1}^{n-1}(A_i\cap A_{n+1})]+P[A_n\cap A_{n+1}]-P\lbrace [\cup_{i=1}^{n-1}(A_i\cap A_{n+1})]\cap [A_n\cap A_{n+1}]\rbrace\nonumber\\
=P[[\cup_{i=1}^{n-1}(A_i\cap A_{n+1})]+P[A_n\cap A_{n+1}]-P\lbrace \cup_{i=1}^{n-1}[(A_i\cap A_{n+1})\cap(A_n\cap A_{n+1})]\rbrace,
\end{eqnarray}
\end{widetext}
where we have used the inclusion-exclusion principle. By using twice the property (\ref{setproperty}) we can rewrite
\begin{widetext}
\begin{eqnarray}
P[\cup_{i=1}^{n-1}(A_i\cap A_{n+1})]+P[A_n\cap A_{n+1}]-P\lbrace \cup_{i=1}^{n-1}[(A_i\cap A_{i+1}\cap...\cap A_n\cap A_{n+1})\cap(A_n\cap A_{n+1})]\rbrace =\nonumber\\
P[\cup_{i=1}^{n-1}(A_i\cap A_{n+1})]+P[A_n\cap A_{n+1}]-P\lbrace \cup_{i=1}^{n-1}(A_i\cap A_{i+1}\cap...\cap A_n\cap A_{n+1})\rbrace=\nonumber\\
P[\cup_{i=1}^{n-1}(A_i\cap A_{n+1})]+P[A_n\cap A_{n+1}]-P\lbrace \cup_{i=1}^n(A_i\cap A_{n+1})\rbrace=\nonumber\\=
P[A_n\cap A_{n+1}],
\end{eqnarray}
\end{widetext}
i.e. our thesis.

We can now prove the theorem 1. We prove it by induction. The theorem holds for $N=2$, because in this case Eq.~(\ref{theorem}) is equivalent to the inclusion-exclusion principle. We assume that Eq.~\ref{theorem} holds for $N$ and we prove that it holds for $N+1$. In fact,
\begin{widetext}
\begin{eqnarray}
P(\cup_{i=1}^{N+1}A_i)=P(\cup_{i=1}^N A_i\cup A_{N+1})=P(\cup_{i=1}^N A_i)+P(A_{N+1})-P[(\cup_{i=1}^N A_i)\cap A_{N+1}]=\nonumber\\
P(\cup_{i=1}^N A_i)+P(A_{N+1})-P[\cup_{i=1}^N (A_i\cap A_{N+1})]=
P(\cup_{i=1}^N A_i)+P(A_{N+1})-P(A_N\cap A_{N+1})=\nonumber \\
\sum_{i=1}^N P(A_i)-\sum_{i=1}^{N-1} P(A_i\cap A_{i+1})+P(A_{N+1})-P(A_N\cap A_{N+1})=\nonumber\\
\sum_{i=1}^{N+1} P(A_i)-\sum_{i=1}^N P(A_i\cap A_{i+1}),
\end{eqnarray}
\end{widetext}
i.e. our thesis. For the benefit of the reader we note that in the second equivalence we use the inclusion-exclusion principle, in the fourth we use the lemma, and in the fifth we use the induction hypothesis, i.e. that the thesis holds for $N$.

In the case of repeats considered in the paper it is $N=\ell-1$ and $P(A_i)=(1-q)^\alpha q^{\ell}$. Moreover for any $i$ it is $P(A_i\cap A_{i+1})=(1-q)^\alpha q^{\ell-1} \tilde q$. From these values and Theorem 1 (i.e. Eq.~\ref{theorem}), Eq.~\ref{IRswithgap2} holds.

\section{Appendix 2}

In this section we derive the expression (\ref{markov_general}) for the expected number of perfect inverted and mirror repeats in a Markovian genome. 

Let us indicate the left part of the stem with $n_1n_2...n_{\ell}$ and consequently the right part of the stem will be $\bar n_{\ell}...\bar n_2\bar n_1$, where the bar indicates matching accordingly to the type of investigated repeats. We shall also indicate with $m_1,...m_m$ the loop and with $x_1$ ($x_2$) the base before (after) the repeat. The repeats can be symbolically expressed as $x_1n_1n_2...n_{\ell}   m_1...m_m \bar n_{\ell}...\bar n_2\bar n_1x_2$. The probability for such a structure is 
\begin{eqnarray}
p(x_1)p(n_1|x_1)p(n_2|n_1)...p(m_1|n_{\ell})p(m_2|m_1).....\nonumber \\
\times p(\bar n_{\ell}|m_m)...p(\bar n_1|\bar n_2) p(x_2|\bar n_1).
\label{markov1}
\end{eqnarray}  
Since we are not interested in the specific bases in $x_1$ and $x_2$ we can sum the probability in Eq. (\ref{markov1}) in $x_1$ and $x_2$ requiring that they are not complementary (remember that we are looking for {\it maximal} repeats). The expression becomes
\begin{eqnarray}
p(n_2|n_1)...p(m_1|n_{\ell})p(m_2|m_1).....p(\bar n_{\ell}|m_m)...p(\bar n_1|\bar n_2)\nonumber\\
\times \sum_{x_1\ne \bar x_2}p(x_1)p(n_1|x_1) p(x_2|\bar n_1).~~~~~~~~~~~
\label{markov2}
\end{eqnarray} 
The sum term in Eq. (\ref{markov2}) becomes
\begin{eqnarray}
\sum_{x_1\ne \bar x_2}p(x_1)p(n_1|x_1) p(x_2|\bar n_1)\nonumber\\
=p(n_1)-\sum_{x=1}^4p(x)p(n_1|x)p(\bar x|\bar n_1),
\end{eqnarray}
where we have used the property $\sum_{x=1}^4 p(x|y)=1$.

In expression \ref{markov1} we need to sum over the possible loop, i.e. in the variables $m_1,...m_m$, by using the constraint $m_1\ne \bar m_m$. We sum first over the internal bases of the loop $m_2,...,m_{m-1}$ obtaining
\begin{widetext}
\begin{equation}
p(m_1|n_{\ell})p(\bar n_{\ell}|m_m)\sum_{m_2,...m_{m-1}} p(m_2|m_1)p(m_3|m_2).....p(m_m|m_{m-1})=p(m_1|n_{\ell})p_{m-1}(m_{m}|m_1)p(\bar n_{\ell}|m_m),
\label{markov3}
\end{equation}
\end{widetext}
where $p_k(b|a)$ is the $k-$step transition probability, i.e. the probability of having the symbol $b$ conditioned to the fact that $k$ step before the symbol was $a$. For Markov chain the $k-$step transition probability matrix is easily obtained as the $k-$th power of the one step transition probability matrix. In obtaining the equation \ref{markov3} we have use the Chapman-Kolmogorov equation, that in its simpler form is $\sum_{z=1}^4 p(y|z)p(z|x)=p_2(y|x)$.

Last we need to sum the expression \ref{markov3} over the variables $m_1$ and $m_m$ by imposing that they are not complementary.  By using again the Chapman-Kolmogorov equation we obtain
\begin{eqnarray}
\sum_{m_1\ne \bar m_m}p(m_1|n_{\ell})p_{m-1}(m_{m}|m_1)p(\bar n_{\ell}|m_m)\nonumber\\
= p_{m+1}(\bar n_{\ell}|n_{\ell})-\sum_{y=1}^4 p(\bar n_{\ell}|y)p_{m-1}(y|\bar y)p(\bar y|n_{\ell}).
\end{eqnarray}
By putting all the terms together we finally obtain
\begin{widetext}
\begin{eqnarray}
\left(p(n_1)-\sum_{x=1}^4p(x)p(n_1|x)p(\bar x|\bar n_1)\right)  p(n_2|n_1)...p(n_{\ell}|n_{\ell-1}) \nonumber \\
\times \left[p_{m+1}(\bar n_{\ell}|n_{\ell})-\sum_{y=1}^4 p(\bar n_{\ell}|y)p_{m-1}(y|\bar y)p(\bar y|n_{\ell})\right] p(\bar n_{\ell-1}|\bar n_{\ell})...p(\bar n_1|\bar n_2) .
\end{eqnarray}
\end{widetext}
that can be simplified by noting that $p(n_1) p(n_2|n_1)...p(n_{\ell}|n_{\ell-1})=p(n_1n_2...n_{\ell})$ is the probability of the $\ell-$word of the left part of the stem. Likewise $p(\bar n_{\ell-1}|\bar n_{\ell})...p(\bar n_1|\bar n_2)=p(\bar n_{\ell}...\bar n_2\bar n_1)/p(\bar n_{\ell})$ is proportional to the probability of the $\ell-$word of the right part of the stem. Hence the probability of a  repeat with a specified sequence in the stem is
\begin{widetext}
\begin{eqnarray}
p(n_1n_2...n_{\ell})p(\bar n_{\ell}...\bar n_2\bar n_1)~\frac{\left(p(n_1)-\sum_{x=1}^4p(x)p(n_1|x)p(\bar x|\bar n_1)\right) \left(p_{m+1}(\bar n_{\ell}|n_{\ell})-\sum_{y=1}^4 p(\bar n_{\ell}|y)p_{m-1}(y|\bar y)p(\bar y|n_{\ell})\right)}{p(n_1)p(\bar n_{\ell})}.
\label{markov}
\end{eqnarray}
\end{widetext}
On the other hand it is easy to see that the corresponding expression for a i.i.d sequence is
\begin{equation}
P_{iid}=p(n_1n_2...n_{\ell})p(\bar n_{\ell}...\bar n_2\bar n_1) (1-\sum_{x=1}^4p(x)p(\bar x))^2.
\label{iid}
\end{equation}
It is direct to show that  Eq.(\ref{markov}) reduces to Eq.(\ref{iid}) when all the transition probabilities satisfy $p(x|y)=p(x)$, i.e. the process has no memory and becomes i.i.d.

In order to obtain the number of  repeats of stem length $\ell$ and loop length $m$ one needs to sum Eq. (\ref{markov}) over the $4^{\ell}$ possible $\ell-$words composing the left part of the stem, i.e.
\begin{widetext} 
\begin{eqnarray}
P_{markov}(\ell,m)=\sum_{n_1,n_2,,...,n_{\ell}=1}^4 p(n_1n_2...n_{\ell})p(\bar n_{\ell}...\bar n_2,\bar n_1)\\~
\times \frac{\left(p(n_1)-\sum_{x=1}^4p(n_1|x)p(\bar x|\bar n_1)\right) \left(p_{m+1}(\bar n_{\ell}|n_{\ell})-\sum_{y=1}^4 p(\bar n_{\ell}|y)p_{m-1}(y|\bar y)p(\bar y|n_{\ell})\right)}{p(n_1)p(\bar n_{\ell})},\nonumber
\end{eqnarray}
\end{widetext}
which is the result of Eq.(\ref{markov_general}).

\end{document}